\documentclass{article}

\usepackage{caption}
\usepackage{subcaption}
\usepackage{arxiv}
\usepackage{graphicx}
\usepackage[utf8]{inputenc} 
\usepackage[T1]{fontenc}    
\usepackage{hyperref}       
\usepackage{url}            
\usepackage{booktabs}       
\usepackage{amsfonts}       
\usepackage{nicefrac}       
\usepackage{microtype}      
\usepackage{lipsum}
\usepackage[symbol]{footmisc}

\title{Explainable AI in Credit Risk Management}

\author{
Branka Hadji Misheva\footnotemark[1] \\
  School of Engineering\\
  Zurich University of Applied Sciences\\
  Winterthur, Switzerland \\
  \texttt{branka.hadjimisheva@zhaw.ch} \\
\And
Ali Hirsa \\
  IEOR Department\\
  Data Science Institute \\
  Columbia University\\
  \texttt{ali.hirsa@columbia.edu} \\
   \And
 Joerg Osterrieder\footnotemark[1] \\
  School of Engineering\\
  Zurich University of Applied Sciences\\
  Winterthur, Switzerland \\
  \texttt{joerg.osterrieder@zhaw.ch} \\
  \And

  Onkar Kulkarni\\
  IEOR Department\\
  Columbia University\\
  New York, NY 10027 \\
  \texttt{ok2294@columbia.edu} \\
   \And
 Stephen Fung Lin \\
  IEOR Department\\
  Columbia University\\
  New York, NY 10027 \\
 \texttt{sfl2122@columbia.edu} \\
}

\begin{document}
\maketitle
\footnotetext[1]{Financial support by the Swiss National Science Foundation within the project “Mathematics and Fintech - the next revolution in the digital transformation of the Finance industry” is gratefully acknowledged by the corresponding author. 
This research has also received funding from the European Union's Horizon 2020 research and innovation program FIN-TECH: A Financial supervision and Technology compliance training programme under the grant agreement No 825215 (Topic: ICT-35-2018, Type of action: CSA).
Moreover, this article is also based upon the work from the Innosuisse Project 41084.1 IP-SBM Towards Explainable Artificial Intelligence and Machine Learning in Credit Risk Management.
Furthermore, this article is based upon work from COST Action 19130 Fintech and Artificial Intelligence in Finance, supported by COST (European Cooperation in Science and Technology), www.cost.eu (Action Chair: Joerg Osterrieder).\newline
The authors are grateful to management committee members of the COST Action CA19130 Fintech and Artificial Intelligence in Finance as well as speakers and participants of the 5th European COST Conference on Artificial Intelligence in Finance and Industry, which took place at Zurich University of Applied Sciences, Switzerland, in September 2020.}

\begin{abstract}

Artificial Intelligence (AI) has created the single biggest technology revolution the world has ever seen. For the finance sector, it provides great opportunities to enhance customer experience, democratize financial services, ensure consumer protection and significantly improve risk management. While it is easier than ever to run state-of-the-art machine learning models, designing and implementing systems that support real-world finance applications have been challenging. In large part because they lack transparency and explainability which are important factors in establishing reliable technology and the research on this topic with a specific focus on applications in credit risk management. In this paper, we implement two advanced post-hoc model agnostic explainability techniques called Local Interpretable Model Agnostic Explanations (LIME) and SHapley Additive exPlanations (SHAP) to ML-based credit scoring models applied to the open-access dataset offered by the US-based P2P Lending Platform, Lending Club. Specifically, we use LIME to explain instances locally and SHAP to get both local and global explanations. We discuss the results in detail and present multiple comparison scenarios by using various kernels available for explaining graphs generated using SHAP values. We also discuss the practical challenges associated with the implementation of these state-of-art eXplainabale AI (XAI) methods and document them for future reference. We have made an effort to document every technical aspect of this research, while at the same time providing a general summary of the conclusions.

\end{abstract}


\section{Introduction}
The management of risks, and especially the management of credit risk is one of the core challenges of financial institutions. Advances in technology have enabled lenders to reduce lending risk by evaluating a profusion of customer data. Using statistical and machine learning (ML) techniques, the available data is analyzed and boiled down to a single quality measure, known as a credit score. Complex machine learning models have been proven very effective in providing a high predictive accuracy  in assessing the credit risk of customers. However, these new and advanced techniques lack the transparency which is necessary to understand how and why a certain borrower is granted a loan or not. Namely, since these models are created directly from data by an algorithm it is often very difficult to trace back the steps the algorithm took to arrive at its decision. Since these models can lack transparency, in some cases significantly, model developers can only interpret the predictions by guessing what the models might have learned. According to  Ribeiro et al. (2016) when we directly use machine learning classifiers as tools for decision making, a vital concern arises: if the users do not trust the model or its predictions, they will not use it. In this context, it is important to differentiate between two different (but related) definitions of trust (\cite{LIME}): 
\begin{itemize}
    \item trusting a prediction, i.e. whether a user trusts an individual prediction sufficiently to take some action based on it,
    \item trusting a model, i.e. whether the user trusts a model to behave in reasonable ways if deployed.
\end{itemize}

This explainability of credit decisions is also explicitly demanded by regulators. A report by the German Federal Financial Supervisory Authority, BAFIN (\cite{BaFin}), from 2018 states that they will not accept any "black box excuses", indicating further that it will be the responsibility of the supervised entity to ensure that any AI-based decisions can be explained to 3rd parties. Furthermore, this issue of explainability is particularly relevant for finance service providers operating in Europe as the General Data Protection Regulation (GDPR) applies across all European Union Member States and Articles 13-15 of this regulation supports a right to an explanation. Specifically, when an individual is subject to a decision based solely on automated processing that significantly affects that individual, the GDPR enables that user the right to meaningful information about the logic involved in the decision taken. This being said, it is clear that having explanations supporting the outputs of a model is crucial and improvements in the understanding of the system can lead to a significant increase in the adoption of innovative methodologies in financial applications. To this end, the concept of eXplainable AI (XAI) emerged introducing a suite of machine learning (ML) techniques that produce models that offer an acceptable trade-off between explainability as well as predictive utility and enables humans to understand, trust and manage the emerging generations of AI models. Among the emerging techniques, two frameworks have been widely recognized as the state-of-the-art in machine learning explainability and those are: 
\begin{itemize}
    \item the LIME framework, introduced by Ribeiro et al. in 2016 (\cite {LIME}) 
    \item SHAP values, introduced by Lundberg et al. in 2017 (\cite {SHAP}).
\end{itemize}

Both methods shed light on the inner workings of black box models, thereby explaining the reasoning behind the predictions. Interpretability is especially important in finance applications where the reliance of the model on the correct features must be guaranteed. Yet, despite the high relevance of this topic in AI's applications in finance, the literature focusing on this specific context has been very limited with only a few exceptions. Bussman et al. (2020) (\cite{BussXAI}) propose a XAI model for fintech risk management based on Shapley values where the authors aim to explain loan decisions using a sample of 15,000 SME companies. The authors confirm the usefulness of Shapley values in increasing the transparency of complex ML models applied to credit risk but they do not tackle some of the main technical challenges in using this approach on larger data sets, which would be the more natural state for finance service providers. The most interesting precedent is perhaps the
research by Miller et al. (2020) (\cite{MillXAI}) where the authors aim to assess the predictive utility of several well-known ML models in the context of P2P lending platforms' credit scoring. Furthermore, the authors apply the Shapley method to provide explainability to the prediction also in view of a much larger feature space. Although this study is one of the few that aims to test the practicality of using emerging XAI techniques in the financial context, it too implements only one of the two main frameworks. The main objective of this paper is to contribute to the existing literature by exploring the utility of both SHAP and LIME frameworks in the context of credit risk management. We further outline the practical hurdles in applying these techniques in financial applications as well as propose solutions to the challenges faced.

\section{Interpretability vs Explainability}
\label{sec:interpretability vs explainability}

There is a fine line between interpretability and explainability. Though sometimes these terms are used interchangeably, they have subtle differences that make it important to understand each concept and comprehend their true meaning. According to Richard Gall (2018) \cite{int vs exp}, "Interpretability is the extent to which cause and effect can be observed within a system. It is about being able to discern the mechanics without necessarily knowing why! On the other hand, explainability is the extent to which the internal mechanics of a ML/DL system can be explained in human terms. Explainability is being able to quite literally explain what is happening inside the black box." (for more details see article \cite{int vs exp}). In the context of this paper, we use LIME to explain model predictions on an instance level, whereas we apply SHAP for global explanations.

\subsection{LIME}
Locally Interpretable Model Agnostic Explanations is a post-hoc model-agnostic explanation technique which aims to approximate any black box machine learning model with a local, interpretable model to explain each individual prediction \cite{LIME}. By model agnostic explanations, the authors suggest that it can be used for explaining any classifier, irrespective of the algorithm used for predictions as LIME is independent of the original classifier. Finally, LIME works locally which in essence means that it is observation specific and similarly as SHAP, will give explanations for every specific observation it has. \\

In terms of the methodology, the way LIME works is that it tries to fit a local model using sample data points that are similar to the instance being explained. The local model can be from the class of potentially interpretable models such as linear models, decision trees, etc. For further details, please see \cite{LIME}. The explanations provided by LIME for each observation $x$ is obtained as follows:
\begin{equation}
\xi (x) = \mathrm{argmin}_{g\in \mathcal{G}} \ L(f, g, \pi _{x}) + \Omega (g) 
\end{equation}

where
$\mathcal{G}$ is the class of potentially interpretable models such as linear models and decision trees,

$\emph{g}\,\in \mathcal{G} $: An explanation considered as a model

$\emph{f}: \mathbb{R}^d  \rightarrow \mathbb{R} $: The main  classifier being explained

$\pi_x (z) $: Proximity measure of an instance  $\emph z$  from  x

$\Omega (\emph g)$: A measure of complexity of the explanation $g \in \mathcal{G}$

The goal is to minimize the locality aware loss \emph L without making  any assumptions about \emph {f}, since a key property of LIME is that it is model agnostic. \emph L is  the measure  of how unfaithful \emph g  is  in approximating \emph f in the locality defined by $\mathcal{\pi} (x)$. 

\begin{figure}[b]
    \centering
    \includegraphics[width=8cm]{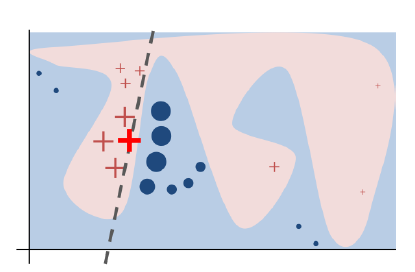}
    \caption{Toy example to present intuition for LIME.}
    \label{fig:fig1}
\end{figure}

\subsection{SHAP}

SHAP, short for SHapley Additive exPlanations, presents a unified framework for interpreting predictions \cite {SHAP}. According to the paper by Lundberg et al. \cite{SHAP}, "SHAP assigns each feature an importance value for a particular prediction. Its novel components include: 
\begin{itemize}
    \item the identification of a new class of additive feature importance measures;
    \item theoretical results showing there is a unique solution in this class with a set of desirable properties.
\end{itemize}

The paper also discusses \emph {local accuracy}, \emph {missingness}, and \emph {consistency}. These are considered as the desirable properties of the unique solution to the class of additive feature attribution methods. The idea is to see the impact each feature has on the target. SHAP values attribute to the change in the expected model prediction when conditioning on that feature \cite{SHAP}. They explain how to get from the base value E[\emph {f(z)}] that would be predicted if the features are unknown to the current output \emph f(x). We do not get into the technicalities of SHAP as it is beyond the scope of this paper. For further understanding of how SHAP works and the theorems developed by the authors, please see \cite{SHAP}.

\begin{figure}[h]
    \centering
    \includegraphics[width=15cm]{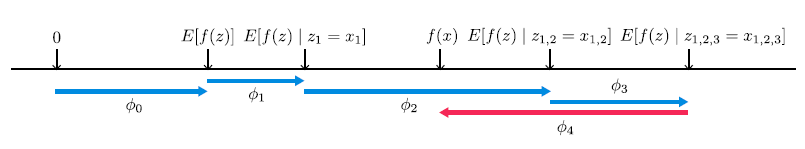}
    \caption{SHAP (SHapley Additive exPlanation) values attribute to each feature the change in the expected model prediction when conditioning on that feature.}
    \label{fig:fig2}
\end{figure}

\section{Data pre-processing and ML models }
\label{sec:ml models}

In this section, we discuss the credit risk data set used in the paper and elaborate on the black box ML/DL classifiers aimed at distinguishing between the different risk classes. We use the Lending Club data set for the experimentation, obtained from Kaggle. It contains records from over 2.2 million peer-to-peer loans issued through the Lending Club platform. All loans had a 3-year term, and the outcome of each loan is known (i.e. we know whether it was fully paid or charged off as a loss). The objective is simple: to build a classifier that has high accuracy in distinguishing between the two classes (i.e. default vs. no default). In order to classify the loans, we train four machine learning based classifiers (i.e. Logistic Regression, XGBoost, Random Forest and Support Vector Machine) and one Neural Network (NN) classifier. 

\subsection{Data}
\label{sec: data}
The original Lending Club data consists of 145 features, ranging from consumer demographic information, to indicators that capture consumer payment behaviour. A sample of the available features can be found in Table \ref{tab:table1}. Based on the available information, our target was to identify the best set of features which result in the highest F1 and Receiver Operating Characteristic (ROC) score on the test data set. The data set available from Kaggle is highly imbalanced. The class of charged off loans (category 1) were very few in number as compared to fully paid loans (category 0). Additionally, there were a lot of empty values in the data, suggesting a very limited usage of the available features. All the data pre-processing steps are explained in Section \ref{sec: preprocess}.

\subsection{Data Pre-Processing}
\label{sec: preprocess}
\begin{itemize}
\item
In order to deal with the missing values, in the first instance, all columns which had "NaN" values in more then 90\% of the records, were canceled. This resulted in a new feature space with 86 features to deal with, and reducing the records to approximately 700,000.

\item
In the next step, highly correlated features were also eliminated from the input space. In order to identify these features, we plotted heat maps depicting the correlation amongst numerical features. In the context of the categorical variables, the selection was made using a chi-square test. 


\item 
For dealing with the categorical variables, we engaged in one hot encoding and in some cases, we combined multiple levels within the categorical feature to reduce the overall data complexity.  For instance, the variable featuring the loan grade ("Grade") was categorized into 6 alphabetical categories from A to G. We reduced the categories to 4 by clubbing E, F and G grade loans into grade D.


\item 
Finally, the target variable which is the loan status ("Status") was converted into a binary categorical variable from the nine categories it was originally broken into. Only two categories were finally kept; "Fully Paid" and "Default". The "Default" category also contained all loans that had the label "Charge-off" in the original classification. 

\end{itemize}

\begin{table}[h]
 \caption{Lending Club Data Dictionary (example features)}
  \centering
  \begin{tabular}{lll}
    \toprule

    Feature     & Description     & Data Type \\
    \midrule
    annualInc & The self-reported
                annual income 
                provided by the
                borrower during
                registration.     &  Numeric  \\
                
    grade     & LC assigned loan 
                grade             & Categorical \\
                
    loanAmnt &  The listed amount
                of the loan applied
                for by the borrower. & Numeric  \\
                
    term     &  The number of
                payments on the
                loan. Values are
                in months and
                can be either
                36 or 60.         & Categorical \\
 
    int\_rate & Interest Rat 
                on the loan       & Numeric \\

    inq\_last\_6mths  & The number
                of inquiries in
                past 6 months
                (excluding auto
                and mortgage 
                inquiries)  & Categorical \\

    total\_pymnt  & Payments 
                received to date
                for total amount
                funded           & Numeric \\

    recoveries  & post charge
                off gross recovery & Numeric \\

    last\_pymnt\_amnt  & Last 
                total payment
                amount received    & Numeric \\
                
    loan\_status  & Current status
                    of the loan 
                    (target variable)   & Categorical \\

    \bottomrule
  \end{tabular}
  \label{tab:table1}
\end{table}

Considering that the main objective of the paper is to implement the two state-of-the-art post-hoc explainability techniques, and outline the main practical challenges when working with financial problem sets, the focus on feature enginering was minimal. Namely, apart from the described pre-processing steps, the majority of the input features were used in their original form. Furture work should focus on extending the pre-processing steps including data-driven feature selection and engineering.


\begin{figure}[h]
    \centering
    \includegraphics[width=8cm]{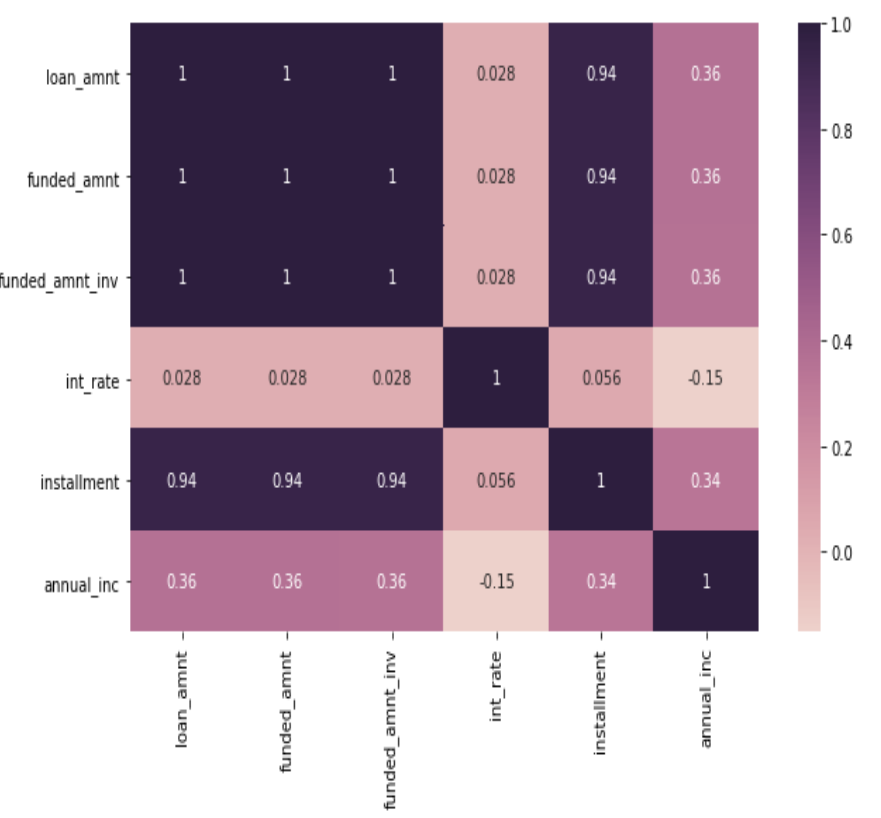}
    \caption{Correlation heatmap amongst numerical features from the Lending Club data.}
    \label{fig:fig3}
\end{figure}

\subsection{Machine Learning Models}
As discussed previously, we train and test four ML-based classifiers and one NN-based binary classifier. Among the ML-classifiers, we trained a very basic yet highly interpretable Logistic Regression Model, followed by the equally widely used tree based models, XGBoost and Random Forest. After iterating over a set of hyper-parameters using grid search and identifying the best set of hyper-parameters, a Support Vector Machine (SVM) Classifier was also developed. A key point to mention with respect to the various ML models tested, is that the SVM classifer yield results quite similar to the other less-computationally intensive methods (i.e. the tree-based classifiers). To expand our experimentation of classification models even further, we developed a Neural Network Classifier. The training time for a simple two-layer architecture trained for 20 epochs was around ten minutes. All the information regarding the hyper-parameter space and model performance metrics on the test set have been elaborated in Table \ref{tab:table2}.   


\begin{table}[h]
 \caption{Model architectures and parameter settings}
  \centering
  \begin{tabular}{lll}
    \toprule
    Model     & Parameter Space     & Performance on Test Data\\
    \midrule
    
    Logistic Regression 
&    
penalty='l2'
solver='lbfgs'

& 
Accuracy: 0.9978 , Precision: 0.9960
\\&& Recall: 0.9932, F1 score: 0.9946
\\
\\
    XGBOOST
&
scoring = 'roc\_auc', cv = 5, n\_jobs = -1, 
& 
Accuracy: 0.9971 , Precision: 1.00

\\

& verbose = 3, n\_estimators = 100,
& Recall: 0.97, F1 score: 0.99
\\
& max\_depths = 4 
&\\
\\
    Random Forest
& 
n\_estimators: 500,
max\_depth: 20
&
Accuracy: 0.9932, Precision: 1.00      
\\
&& 
Recall: 0.96, F1 score: 0.98 \\
\\
\\
    SVM
&
gamma='auto',
C=1.0, kernel='rbf', 
&
Accuracy: 0.99487, Precision: 1.00           

\\
&
probability=False/True
& 
Recall: 0.96, F1 score: 0.98
\\
\\

    Neural Networks
&
n\_hidden = 2, neurons = [35,35], 
&
Accuracy: 0.9998, Precision: 0.9999

\\ 
& activations = RelU, sigmoid 
&
Recall: 0.9985, F1 score: 0.9992

\\
& loss = binary\_crossentropy , Optimizer = adam 
&

  \end{tabular}
  \label{tab:table2}
\end{table}

\section{Using LIME to explain Local Instances}

For obtaining the instance explanations using the LIME framework, the lime explainers available in the Python library "LIME" were used. For our specific use case, we used the LIME implementation for tabular data and the specific steps involved in getting the model explanations are:

\begin{itemize}

\item
Create a list of all the features names.
\item
Create a list of Target label categories, i.e. "Fully Paid" and "Default".
\item
Create a function that takes in an array of test instances (feature values) and returns class probabilities.
\item
In the lime tabular explainer object, pass the training data, features list, class list and the probability returning function mentioned above.
\item
After creating the explainer, select an instance to be explained and pass it as an argument to the explainer. The output is then a list of top 10 features affecting the model's prediction for that instance.
\end{itemize}

\begin{figure}[h]
    \centering
    \includegraphics[width=15cm]{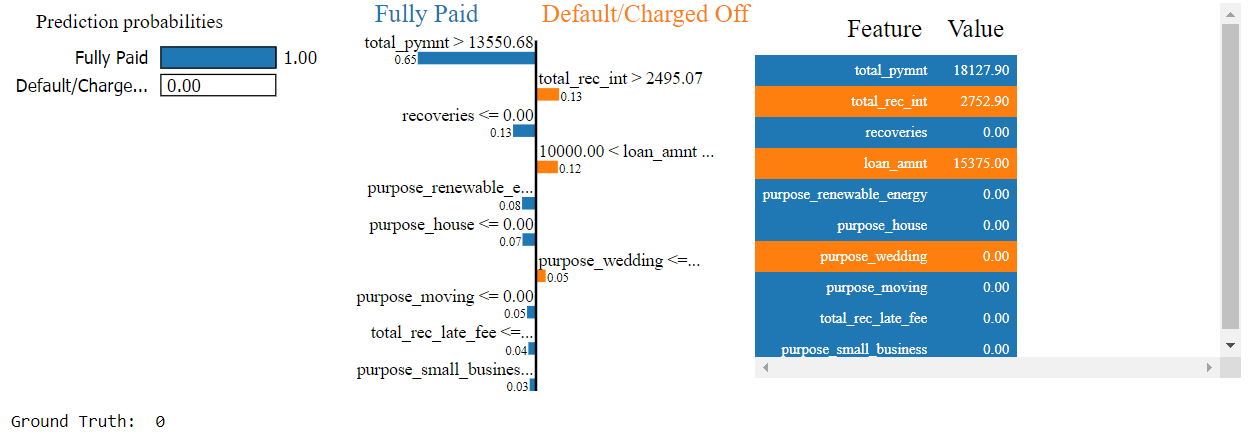}
    \caption{LIME explanation for a customer classified as a "Fully Paid" Loan type by XGBOOST Model.}
    \label{fig:fig4}
\end{figure}

The visualization of the main features that have contributed towards pushing the probabilities towards either class are represented in Figure \ref{fig:fig4}.


\subsection{LIME on Tree Based Models}

After establishing the steps involved in getting the explanations from the LIME implementation, we attempted to apply the procedure to the trained XGBOOST classifier. The main challenge we faced in this context is associated with the implementation of the LIME framework in Python which in turn cannot accommodate models with GridSearchCV objects. As a result, we re-train the model using the best parameter set obtained during the hyper-parameter grid search which in turn ensured that the model is stored as an xgboost object.


\begin{figure}[h]
    \centering
    \includegraphics[width=15cm]{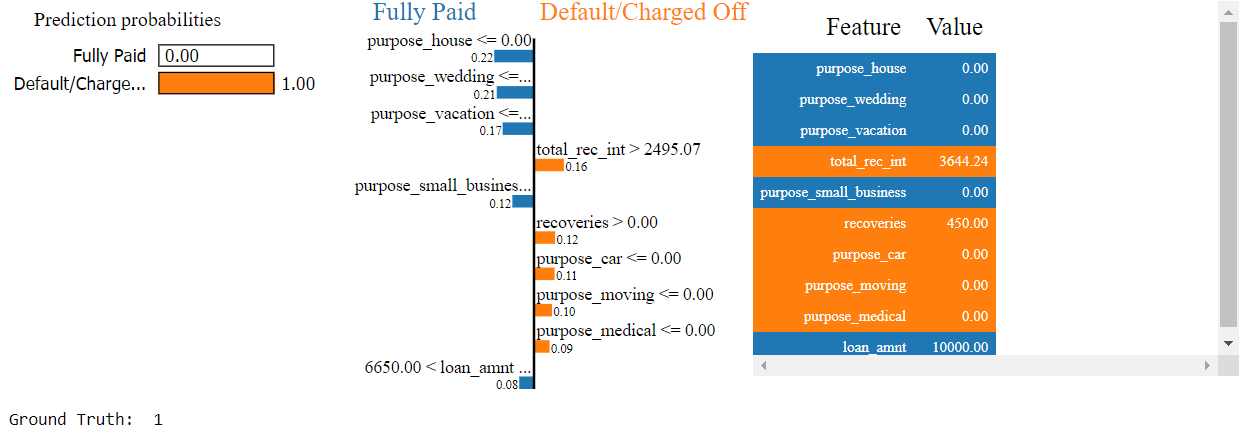}
    \caption{LIME explanation for a customer classified as a "Default" Loan type by XGBOOST Model.}
    \label{fig:fig5}
\end{figure}

\begin{figure}[t]
    \centering
    \includegraphics[width=15cm]{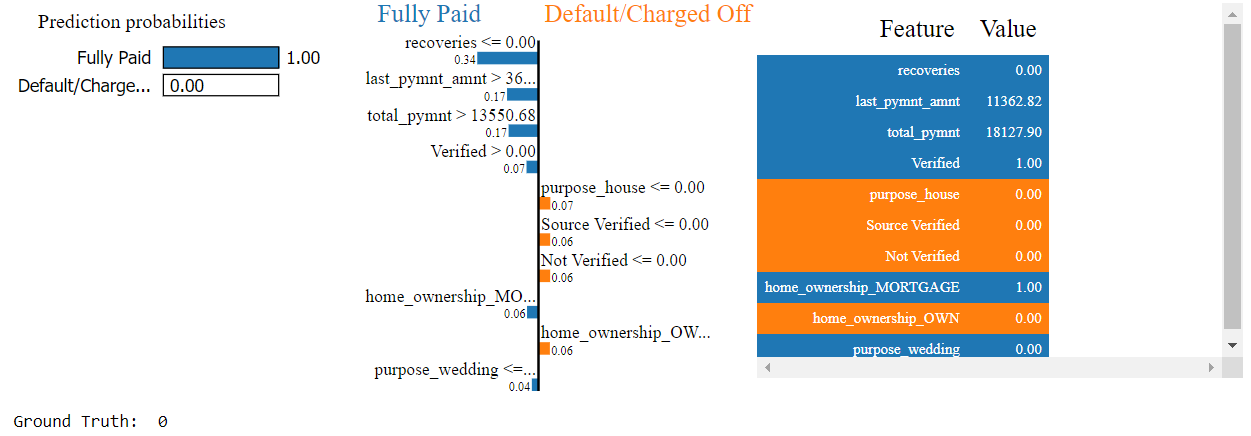}
    \caption{LIME explanation for a customer classified as a "Default" Loan type by Random Forest Model.}
    \label{fig:fig6}
\end{figure}


\subsubsection{Interpretation}

 In this section, with the purpose of expanding further on the explanations provided by the LIME framework, we interpret three instances as predicted by the model. In Figure \ref{fig:fig4}, we present an example of a loan contract for which the model has predicted a class "0" output (i.e. the loan contract will not default).  
 Figure \ref{fig:fig4} furthermore shows the top ten features that contributed to this decision along with their contributions. On the left is the model confidence about its prediction. We can interpret this as follows:

\begin{itemize}
\item
Since the value of the "total\_payment" variable was \$18,127.90, which is greater than \$13,550.68 (a value used by the model for making a decision), this pushed the prediction towards the "Fully Paid" category.

\item
Similarly, since there were no recoveries made on the loan, the model pushes the prediction towards the "Fully Paid" category

\item
The "loan\_amnt" > \$10,000 pushes the prediction towards "Default" however, the weight for this feature is not very high. Similarly since "total\_rec\_int" is high, the model pushes the prediction towards "Default".

\item
Overall, the model predicts with absolute certainty that given the feature values for this particular customer, the loan contract will not default. Furthermore, by using the LIME framework, the model developers can also provide the reasoning behind the decision taken to the end users.   

\end{itemize}

In Figure \ref{fig:fig5}, we present our second example of a loan contract for which the model has predicted a class "1" output (i.e. the loan contract will default). Figure \ref{fig:fig5}, similar to Figure \ref{fig:fig4}, illustrates the top ten features that contributed to this decision along with their respective contributions. On the left is the model confidence about its prediction. We can interpret this as follows:

\begin{itemize}
\item
This customer has been categorized as "Default", meaning that they are expected to fail to fully repay the loan. The fact that the purpose of the loan was not for "house", "wedding" or "vacation" pushed the prediction towards the "Fully Paid" category.

\item
However, since the total interest received was \$3,644.24, greater than \$2,495.07, this difference pushes the prediction towards "Default" and the impact it has is higher than many other features. Similarly, since the recoveries made on the account were positive (\$450), this pushed the prediction further towards the "Default" category.

\item
An important observation to note here is that even though the total tally of the top ten features is in favor of pushing the prediction towards "Fully Paid" category, the total tally of all the features combined will be in favor of the "Default" category.

\end{itemize}

In Figure \ref{fig:fig6}, we present our final example of a loan contract for which the Random Forest model has predicted a class "0" output (i.e. the loan contract will not default). Figure \ref{fig:fig6} further shows the top ten features that contributed to this decision along with their respective contributions. On the left is the model confidence about its prediction. We can interpret this as follows:

\begin{itemize}
\item
This customer has been categorized as "Fully Paid", implying that they are expected to fully repay the loan. The fact that no recoveries were made on this account and the last payment amount was large (\$11,362.82 which is by far greater than \$3,600), pushes the prediction towards the "Fully paid" category.

\item
Another important feature of this loan that pushes it towards the risk free category is the fact that it is a "verified" loan. Lending Club verifies a loan before approval and since it was verified by LC, it pushed the prediction further towards the risk free category. Similarly, the total payments received from the account is \$18,127.90 which is higher than the threshold set by the model for decision making and hence pushes it further ahead.

\item
The fact that the loan applicant is a home owner pushes the prediction towards the "Default" category which is somewhat contrary to conventional logic. Nonetheless, this is not enough to turn the prediction around. 
\end{itemize}





After going through all the interpretations from the tree-based models (i.e. XGBOOST and Random Forest), we can say that the reasoning given by the model for making a prediction is in line with financial logic. Ideally, a consumer who has no recoveries against his/her account, no late fees, lower interest rate and higher total payments, can be considered as a risk free customer and hence, companies can have confidence while approving their loans. This was evident from the examples we discussed above. Similarly, high risk customers can be identified by the virtue of their account i.e. late fees, lower repayment amounts, some recoveries made and not verified loans. This is extremely helpful information which helps both model developers and end users increase their trust in the predictions. Furthermore, if model developers observe some discrepancies in model predictions, they can run LIME and understand which features are causing trouble and then decide which features to keep or drop from follow-up classifiers.


\subsection{LIME on SVM}
\label{sec:LIME-SVM}

According to Vapnik et al.\cite{SVM} the way a support-vector classifier works is, it constructs a hyperplane or set of hyperplanes in a high-dimensional space, which can be used for classification. Intuitively, a good separation is achieved by the hyperplane that has the largest distance to the nearest training-data point of any class, since in general the larger the margin, the lower the generalization error of the classifier. An SVM training algorithm builds a model that assigns new examples to one category or the other, making it a non-probabilistic binary linear classifier\cite{skl}. 

We elaborate on two technical solutions for generating probabilities from the SVM model output:

\begin{itemize}
\item

The first method to get the probabilities out of the SVM model is to set the "probability" argument to "True"\cite{skl1}. This assures that we can use the "predict\_proba" function available in scikit-learn and get prediction probabilities.

\item

The second way to generate probabilities out of the SVM model output is to pass all the output through a "sigmoid" function and generate probabilities. The output from SVM's "decision\_function" is the distance of that instance from the separating hyper-plane. The signs help us in understanding which side of the hyperplane does a prediction fall. As we pass these distances through a softmax function, they are converted into probabilities which can further be used for executing LIME. 

\end{itemize}

In terms of model performance, our results suggest that the SVM classifier's predictive utility is significantly lower compared to the other classifiers considered. The most likely reason for such an outcome is overfitting. Furthermore, in the context of LIME explanations, in testing both approaches listed above, we find that only the first method allows for the LIME explainer to compute successfully. Regardless of this, in our unique context, the explanations were not relevant as the underlining model has very poor predictive performance.


\begin{figure}[t]
    \centering
    \includegraphics[width=15cm]{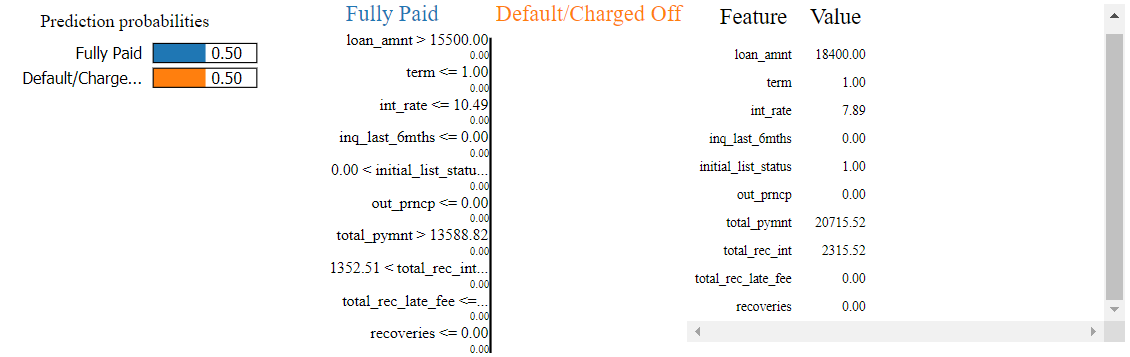}
    \caption{LIME explanation for a customer not clearly classified by the SVM Model.}
    \label{fig:fig8}
\end{figure}

\section{Using SHAP for global explanations}
In this section, we report the results from the implementation of the second state-of-the-art post-hoc explainability technique, SHAP. We visualize the mean SHAP values for a set of data inputs for which explanations of importance and impact are required. The key difference between SHAP and feature importance scores from tree based models is that along with the level of importance of a feature for the model, SHAP also helps us understand the effect each feature has on the prediction. To elaborate on this point in further detail, we present the visual representation of the SHAP values in Figure \ref{fig:fig9}. By looking at the visualization, we can identify the main features that drive the output (features are arranged in descending order of importance hence the most important variables are: (1) "total\_pymnt", (2), "loan\_amount", (3) "least\_pymnt\_amount" etc.) and we can identify the impact that each feature has on the model's output (for example, the explanations provided suggest that the higher the values of the "total\_pymnt" variable, the lower the probability of default). All the interpretations concerning the SHAP values together with the challenges faced while implementing the technique, have been documented in the next sections.


\subsection{SHAP on Tree Based Models}

\begin{figure}[h]
\centering
\begin{subfigure}{.5\textwidth}
  \centering
  \includegraphics[width=1\linewidth]{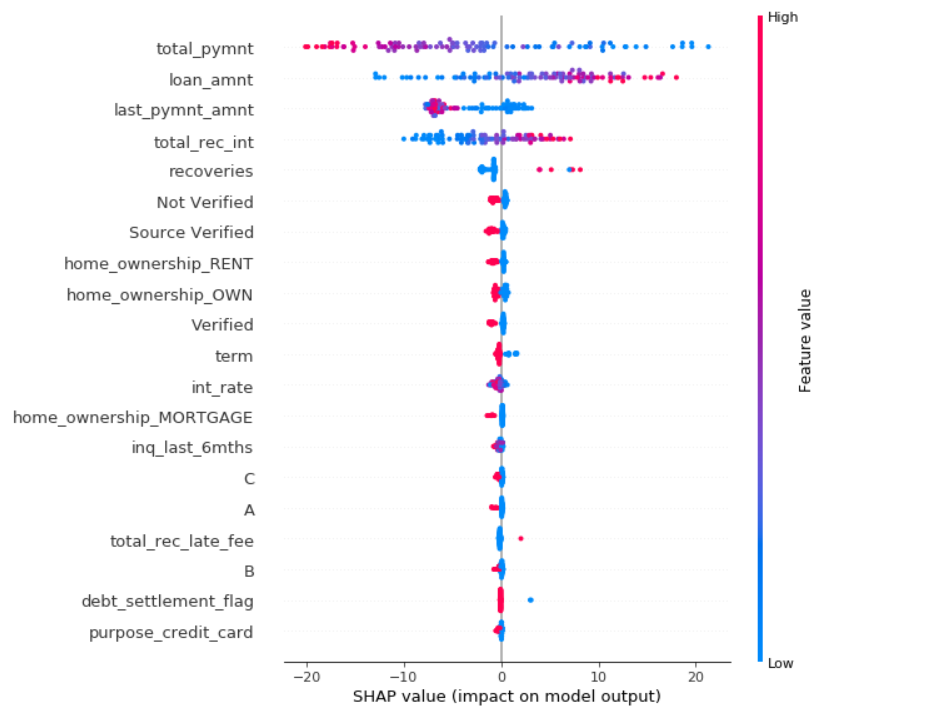}
  \caption{Summary Plot - XGBOOST model using Tree Explainer.}
  \label{fig:sub1}
\end{subfigure}%
\begin{subfigure}{.5\textwidth}
  \centering
  \includegraphics[width=0.9\linewidth]{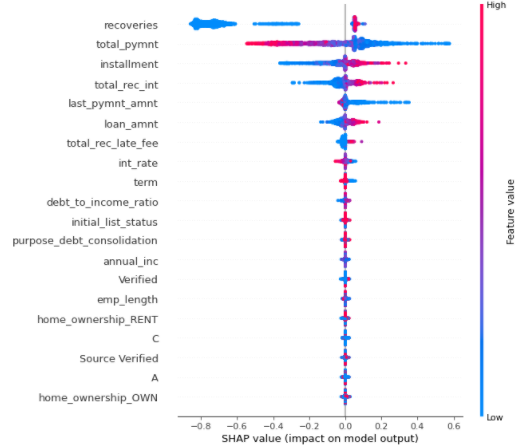}
  \caption{Summary Plot - XGBOOST model using Kernel Explainer.}
  \label{fig:sub2}
\end{subfigure}
\caption{SHAP Kernel Comparison Plots explaining global impact of each feature on XGBOOST model output}
\label{fig:fig9}
\end{figure}

SHAP library allows for model-specific explainers. For instance, to explain tree based models like Random Forest and XGBOOST, tree explainers can be used to generate the mean SHAP values. For experimentation purposes, we tested the various explainers available among which are the tree, kernel, linear and deep explainer. The global impact of the top 20 features on the model's prediction have been explained in Figure \ref{fig:fig9} 


\begin{figure}[b]
    \centering
    \includegraphics[width=10cm]{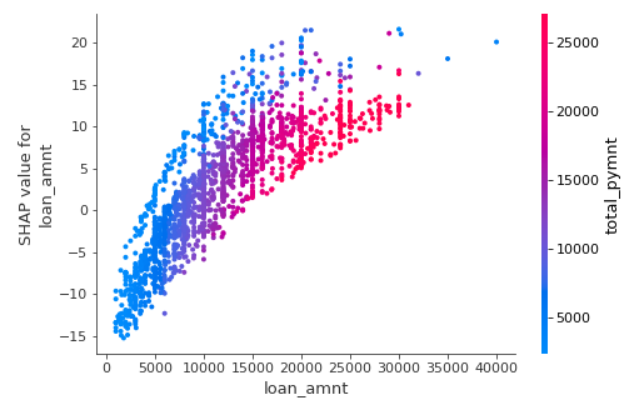}
    \caption{SHAP Dependence Plot.}
    \label{fig:fig10}
\end{figure}

\begin{figure}[t]
    \centering
    \includegraphics[width=15cm]{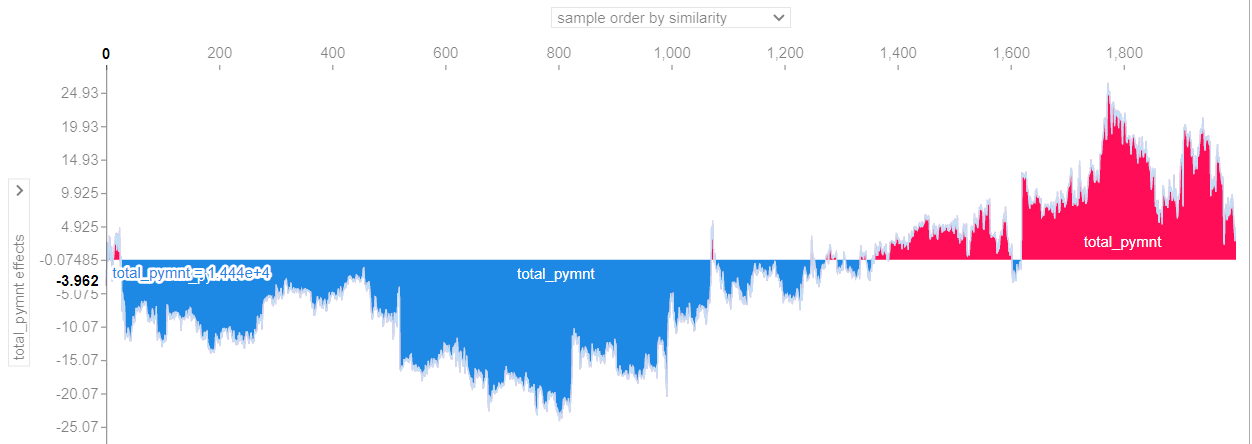}
    \caption{SHAP Force Plot.}
    \label{fig:fig11}
\end{figure}

\subsubsection{Interpretations}

Example 1, see Figure \ref{fig:fig9}. Summary plots are used to explain the global impact of the top 20 features on the model output. This graph informs us not only of the importance of each feature, but also on the impact it has on the model's output. In Figure \ref{fig:sub1}, we generated SHAP values using tree explainers for a sample of 100 test data points. Our initial goal was to generate SHAP values across the entire test data of 100,000 plus data points, however, due to the exponential time complexity of SHAP kernel explainers, we were not able to generate SHAP values for the entire data set. As a result, we generate these values only for a small sample of the data. We experimented with a sample of 100 test data points and 2,000 data points. Figure \ref{fig:sub2} demonstrates the summary plot for explaining the XGBOOST's output using the kernel explainer on 2,000 test data points. These interpretations can be made from Figure \ref{fig:fig9}:

\begin{itemize}
\item
"total\_pymnt" , "loan\_amnt", "last\_pymnt\_amnt", "total\_rec\_int" and "recoveries" are the top five important features. 

\item
The SHAP values suggest that higher "total\_pymnt" amounts (red colored dots) are associated with a lower probability of default whereas lower "total\_pymnt" (blue colored dots) are associated with a higher probability of default. The association with the target variable is inferred from the x-axis with positive values implying the "Default" category and negative values implying the "Fully Paid" category.

\item
Similarly, lower or no recoveries are associated with a lower probability of default whether as higher recoveries push the probability of default upwards.  

\item
In the context of home ownership, the SHAP values suggest that owning a house does not have a significant impact on the model's output. 

\end{itemize}

One key observation from Figure \ref{fig:fig9} is that both explainers are consistent in terms of the top 20 most relevant features and their expected impact on the model's output even though the explanations are summarized across very different test data sets (100 vs. 2,000 data points for Figure \ref{fig:fig9}a and Figure \ref{fig:fig9}b, respectively). This indicates that explanations remain consistent which in turn increases our confidence in the model. Yet another observation from Figure \ref{fig:fig9} is that not all the features are monotonic. That is, the color does not slide uniformly from red to blue hence further research can focus on investigating further these non-linear associations and provide a possible financial rational.



Example 2, see Figure \ref{fig:fig10}. The SHAP Library also provides a dependence plot for analysing the interactions between features as well as their association with the target variable. The idea is very similar to a partial dependence plot and thus the interpretation is similar as well. 



\begin{itemize}
\item

In Figure \ref{fig:fig10}, on the x-axis the value the feature "loan\_amnt" is represented, and on the Y-axis the SHAP value (i.e. the impact it has on the model output) is represented. As discussed previously, a positive impact means the variable is pushing the model output towards the "Default" class and a negative impact means the prediction is pushed towards the "Fully Paid" class. Hence, looking at the "loan\_amnt" dependence plot, we can infer that lower loan amounts are associated with the fully repay class whereas, as the loan amount goes high, the probability of default increases.


\item
The dependence plot in SHAP also represents the interactions between input features. Namely, the plot picks a second feature, which has the highest interaction with the first variable we are investigating and colors the instances according to the value of the second feature \cite{kdn}. In the case of the "loan\_amnt" feature, as a second variable, the model has picked "total\_amnt" and we can see that most of the dispersion we find in the "loan\_amnt" axis is explained by the total payment made.


\end{itemize}

Example 3, See Figure \ref{fig:fig11}. The force plots from SHAP allow us to give explainability to each single model prediction. Specifically, if we take many instance level explanations, rotate them 90 degrees, and then stack them horizontally, we can see explanations for an entire data set (in this case, 2,000 data points).

\begin{itemize}
\item
We include the force plot for one feature: the "total\_pymnt". The numbers on top of the X-axis are the number of instances explained, whereas the Y-axis is the "total\_pymnt" effects on the model prediction. The dependence represented can be interpreted as: the higher the total payment amount, the lower the likelihood of being tagged as defaulting loan and vice versa.

\item
We can also change the y-axis and select any feature whose global effects we wish to assess. The selection "\emph{f(x)}" clubs together the explanations of all the instances and the impact of all the features together, thus providing a holistic view across all the data points.

    
\end{itemize}

\subsection{SHAP on Neural Networks}

\subsubsection{Model Development}

To develop the NN classifier, we conducted several experiments. We developed various model architectures by changing the number of hidden layers and the number of neurons in each hidden layer. We varied the learning rate and also tried changing the activation functions between "tanh" and "relu". Since the problem at hand was a binary classification, the "binary cross entropy loss" function was used. After multiple trials, we selected the model with the best ROC\_AUC score on the test set. The final model parameters can be found in Table \ref{tab:table2}. 

\subsubsection{Explanations}
We used the deep explainer and kernel explainer for explaining the Neural Network classifier. As mentioned before, we experimented with 100, 1,000 and 2,000 samples of test data and present the interpretations drawn from these samples. As we began with the deep explainer, we came across subtle differences from tree and kernel explainers. For instance, Deep explainers provided the SHAP values combined for both the classes, whereas in case of the tree explainers, the SHAP values were provided for all the features by class (i.e. SHAP values are given out separately for each class). However, these values are exactly opposite of each other and hence, explaining one would be sufficient. Additionally, we also observed that deep explainer had a lower computation time in comparison to kernel explainer. With this we are able to conclude that using model specific explainers leads to much faster computations as they are built to accommodate the specific model architectures.

\begin{figure}[t]
\centering
\begin{subfigure}{.5\textwidth}
  \centering
  \includegraphics[width=0.9\linewidth]{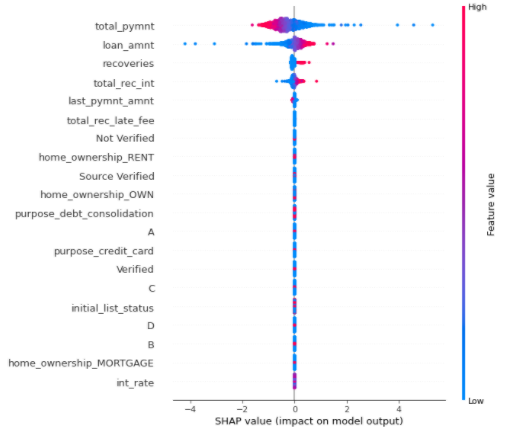}
  \caption{Summary Plot - NN model using Deep Explainer.}
  \label{fig:sub3}
\end{subfigure}%
\begin{subfigure}{.5\textwidth}
  \centering
  \includegraphics[width=0.9\linewidth]{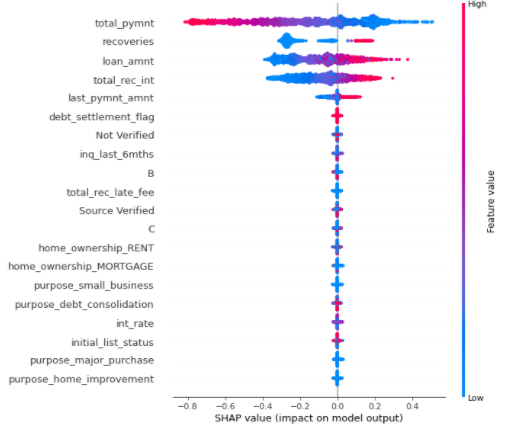}
  \caption{Summary Plot - NN  model using Kernel Explainer.}
  \label{fig:sub4}
\end{subfigure}
\caption{SHAP Kernel Comparison Plots explaining global impact of each feature on NN model output on 2,000 test samples}
\label{fig:fig12}
\end{figure}

\begin{figure}[b]
    \centering
    \includegraphics[width=10cm]{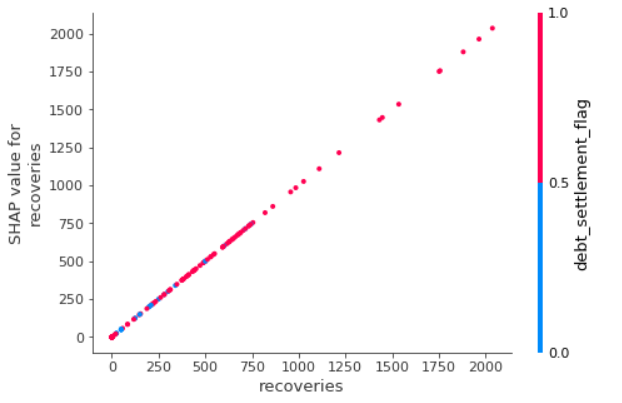}
    \caption{SHAP NN Dependence Plot.}
    \label{fig:fig13}
\end{figure}

\subsubsection{Interpretations}

Example 1, See Figure \ref{fig:fig12}. We compare summary plots from the deep and kernel explainers which ran on 2,000 test data points. Interestingly, there is a considerable overlap amongst the top 20 important features returned by both explainers with only the positioning of certain features being slightly different. Some further observations:

\begin{itemize}


\item
The interpretations are in line with what we observed from the tree explainers, suggesting that lower loan amounts are associated with a lower probability of default. Similarly, positive recoveries made on the account imply high risk customers. 

\item
Some notable differences between the deep and kernel explainer can be observed in features like "loan\_amnt" and "total\_rec\_int". In the case of the kernel explainer, the instances show a smoother transition, implying that there can be some customers who have higher loan amounts and yet be low risk. Likewise for the total interest received, the interpretations are clearer in case of kernel explainer, i.e. lower interest received implies higher likelihood of being tagged as risk free consumers. 

\end{itemize}

Example 2, See Figure \ref{fig:fig13}. In case of the dependence plots, one key observation is that there is not much interaction amongst the features and hence there is no vertical dispersion of the variables being explained (in the specific context: "recoveries" and the effect of debt\_settlement\_flag). That is the reason why we see a straight line with almost a 45 degree slope, implying no interactions of the debt settle flag with the recoveries feature.



\subsection{SHAP Global Explanations vs Feature Importance Scores}

\begin{figure}[t]
\centering
\begin{subfigure}{.5\textwidth}
  \centering
  \includegraphics[width=1\linewidth]{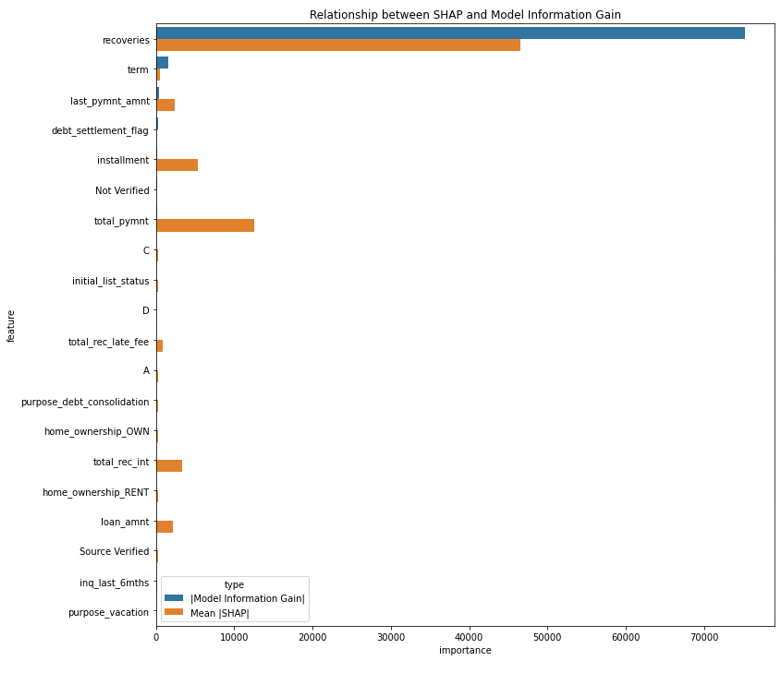}
  \caption{XGBOOST Information Gain vs SHAP Global Explanations.}
  \label{fig:sub5}
\end{subfigure}%
\begin{subfigure}{.5\textwidth}
  \centering
  \includegraphics[width=1\linewidth]{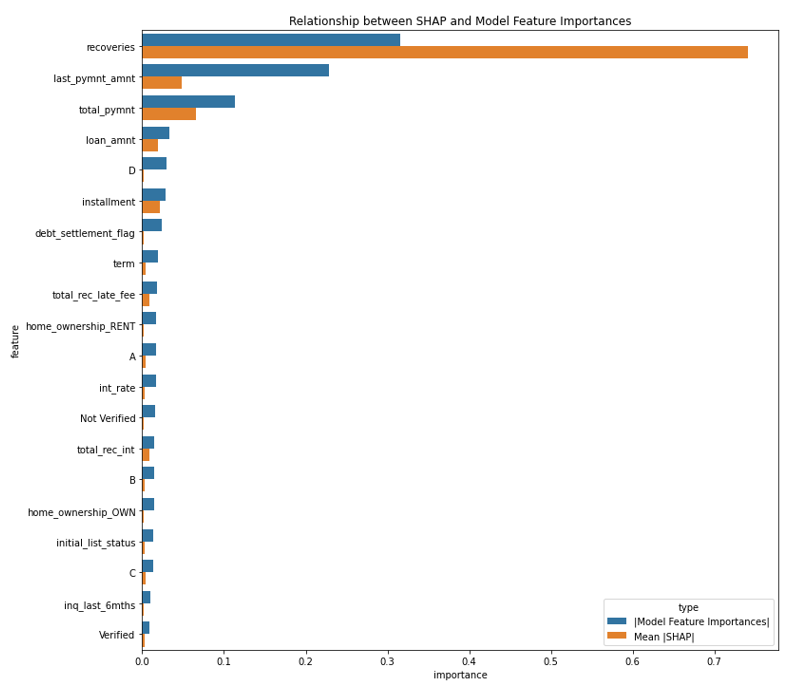}
  \caption{RF Information Gain vs SHAP Global Explanations.}
  \label{fig:sub6}
\end{subfigure}
\caption{Information Gain vs SHAP Global Explanations for XGBOOST and Random Forest Models}
\label{fig:fig14}
\end{figure}

In order to further enhance our understanding of interpretability and explainability, we also conducted experiments to compare the feature importance scores from each model with their respective SHAP global explanations. Even though the comparison may not be fair given that feature importance scores are drawn from the entire training set whereas SHAP global explanations are drawn from a sample of 2,000 test data points, we undertook this study to understand if these two methods convey the same information and ranking among features. Figure \ref{fig:fig14}, presents a comparison between the feature importance scores calculated using information gain and the SHAP mean feature impact. We can clearly observe that different features contribute to the SHAP mean absolute values (spread across the top five) whereas the only significant feature drawn by the information gain approach, is "recoveries". 

\subsection{Accumulated Local Effects}

In order to further confirm our results, we run an additional explanainability technique - the accumulated local effects (ALE) plot. ALE plots are a faster and unbiased alternative to partial dependence plots \cite{book}. Furthermore, the computation of a partial dependence plot for a feature that is strongly correlated with other features involves averaging predictions of data instances that are unlikely in reality. Piece-wise constant models such as XGBoost and random forests do not have continuous prediction functions, which implies they are not differentiable. Thus, we computed ALE plots only for the Logistic Regression, SVM, and Neural Network classifiers. The main challenge for this part is choosing the interval size. Namely, in order to get stable estimations for the ALE plot, we must have a sufficient number of data points within the interval. On the other hand, another crucial determinant of a good estimation to have small enough intervals especially in regions where the prediction function is far from linear with respect to the feature of interest. Therefore, to the best of our knowledge, there is no optimal solution for setting the number of intervals. In the context of this work, we set the interval size by visualization so that the plots contain a similar number of data points in each interval. Additionally, we tried different combinations (increasing the number of intervals) in order to confirm that the plots remain relatively unchanged.

\begin{figure}[b]
    \centering
    \includegraphics[width=15cm]{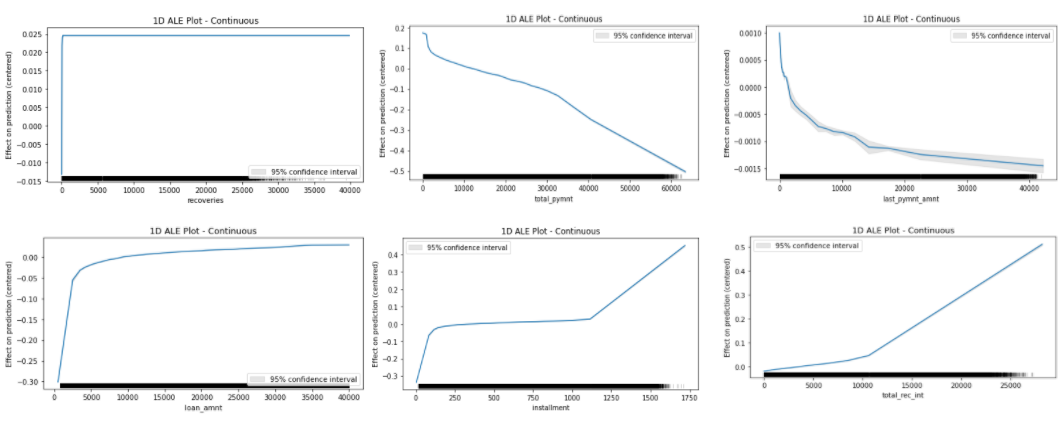}
    \caption{ALE plots - Logistic Regression}
    \label{fig:fig15}
\end{figure}

\subsubsection{Interpretations}

\begin{itemize}
\item 
Example 1, See Figure \ref{fig:fig15}. In the top feature "total\_pymnt" the plot shows that the value of the prediction decreases by approximately 0.65 points (we are predicting a 0 or 1) when the total\_pymnt increases from 0 to 40,000. This indicates that the prediction is being pushed towards the "Fully Paid" category, implying that as the total payment increases, the chances of being tagged as risk free consumers are higher.  

\item

Similarly, as the total received interest goes up, the chances of being tagged as a high risk consumer increases.
\end{itemize}

It is important to note that the y-axis indicates the magnitude by which each feature impacts the prediction hence we might also say that the volume of total payments has a higher impact on the model prediction compared to the total interest received.


\section{Practical Challenges}

This sections outlines some of the main challenges we faced in implementing the post-hoc explainability techniques on financial problem sets as well as the proposed solutions.


\begin{itemize}

\item 
\textbf{Running LIME on the GridSearchCV model object}: To tackle this, we selected the best model hyper-parameters and trained an independent model. We then implemented the LIME explainer.

\item 
\textbf{Generating probabilities from SVM} : We tried two methods for generating probabilities as per LIME's requirements. See section \ref{sec:LIME-SVM}

\item 
\textbf{Identifying differences in SHAP outputs from various explainers}: We had to understand which explainers returned combined SHAP values for both classes and which did not, and generate graphs accordingly.

\item 
\textbf{Longer execution time for Kernel explainer}: We tried generating SHAP values for the entire data set with significant computing power requirements. Hence, we opted for generating SHAP values for a sample of test data points. Since kernel explainer computations have exponential computing requirements, we chose a 2000-row sub-sample from our testing set, to speed up the time-consuming SHAP calculations.

\item 
\textbf{Speeding up SHAP values generation}:
On the background data, it is slow to use the entire training data set, so we used shap.kmeans to create 30 representative rows from another sub-sample of 20,000 points from the original training set.
\end{itemize}

\section{Conclusion and Future Work}
The lack of algorithmic transparency is one of the main barriers for the wider adoption of AI-based solutions in credit risk management. The greater the trust in AI, the more loan originators will deploy it; which in turn will enable them to foster innovation and move ahead in adopting next generation capabilities. From an aggregate perspective, a wide adoption of AI-based solutions in credit risk management may lead to broad benefits for financial inclusion and financial system diversity. By motivating loan originators to fully utilize AI and ML in their credit scoring tasks, they would be able to move some borrowers who would have been classified as subprime by traditional criteria to “better” loan grades, which in turn allows them to get lower priced credit, thus improving financial inclusion. Additional to this, the wide adoption of AI-based solutions could lead to financial system diversity which in turn means lower risks that the economy faces compared to when a few banks dominate credit provision. The main aim of this paper is to implement two advanced post-hoc explainability techniques (LIME and SHAP) on outputs obtained from ML-based credit scoring models. Our results indicate that both LIME and SHAP provide consistent explanations that are in line with financial logic. Furthermore, the 20 most important features remained stable even in view of changing test sizes which in turn only increases our confidence in the conclusions. The results of this work notwithstanding, there is still significant potential in developing robust and reliable XAI techniques in the ML industry as we need to make dedicated efforts to develop XAI techniques that take into account the practical constraints involved in using these methods. Specifically, SHAP values are robust and effectively communicate the importance each feature has over the model prediction. However, in the case of many features, it can take an extremely long time to generate these values, owing to its exponential run time. Similarly, on the other hand, LIME has certain limitations on model objects and the types of model that it can explain (probabilistic models only).

\bibliographystyle{unsrt}  




\end{document}